\documentclass{ws-mpla}
\usepackage[super]{cite}
\usepackage{cancel}
\usepackage[normalem]{ulem}
\usepackage{xcolor}
\usepackage{color,soul}
\usepackage{graphicx}
\usepackage{subcaption}
\usepackage{hyperref} 
\begin{document}

\markboth{D. Barbu et al.}{Nuclear effects in proton decay}

\catchline{}{}{}{}{}

\title{Nuclear effects in proton decay}

\author{Denis Barbu$^1$, Mihaela Parvu$^{1,}$\footnote{Corresponding author: mihaela.parvu@unibuc.ro} , Ionel Lazanu$^1$}

\address{$^1$Faculty of Physics, University of Bucharest, Atomistilor 405, RO-077125, POB-MG11, M\u{a}gurele-Bucharest,
Romania\\
denis.barbu@g.unibuc.ro
}

\maketitle


\begin{abstract}
Proton decay detection could put in evidence physics beyond the Standard Model (BSM). In this context, multiple projects are searching for such events. We focused our work on two of the expected decay modes, $p \rightarrow \pi^+ + \bar{\nu}$ and $p \rightarrow K^+ + \bar{\nu}$. Neutrinos are particles that interact very weakly with matter, so they are not of interest in our work. The detector materials investigated in this study include liquid argon (LAr), liquid xenon (LXe), and water (H$_2$O). Our analysis focuses on two key effects relevant to this decay process: the Fermi motion of nucleons and final state interactions. In addition to these effects, we have also examined the nuclear interaction of particles with the nuclei of the medium. This paper presents values for nucleon energy distribution, cross sections, mean free paths, and interaction probabilities.

\end{abstract}


\section{Historical aspects of the conservation of baryonic number and proton decay}

In 1929, Weyl \cite{weyl1929electron} first discusses the necessity of conserving the number of protons in nuclear processes. This led to the subsequent emergence of two distinct conservation laws: separate conservation of the number of electrons and protons, the first step of the gauge invariance. In nuclear physics, the conservation of the number of nucleons in nuclear processes is a fundamental principle \cite{reines1954conservation}, stemming from the work of St\"ueckelberg, who introduced the concept of baryonic number \cite{stuckelberg1938interaction}. This conservation law dictates the stability of the proton as a particle.
In the Standard Model (SM), the conservation of baryon (B) and lepton (L) numbers is upheld by gauge invariance and renormalization, which guarantee their status as global symmetries within the theory. At the non-renormalization level, the violation of B and L suppose two distinct possibilities: the gauge group of the SM is extended, or the particle content of the SM is supplemented with other particles.
The concept of a finite lifetime for the proton was initially proposed by A. Sakharov \cite{Sakharov:1967dj}. He invoked this notion to explain the matter-antimatter asymmetry within the Big Bang model, but his idea requires certain theoretical justifications. His predictions led to an extremely long lifetime, of about $10^{50}$ years.
In accord with PDG \cite{Workman:2022ynf}, the current limits for the partial mean life as maximal value, is $>1.6\times10^{34}$ years (CL=90\%) predicted for the $p \to \pi^0+\pi^+$ channel.
This paper is a phenomenological study that is focused on experimental aspects of interest at the moment.

\section{Proton decay channels}

Depending on the hypotheses considered, a multitude of decay channels with diverse characteristics are possible. Each proton decay channel can reveal various aspects of the physics beyond the Standard Model (BSM) \cite{nagata2014sfermion}.

\begin{itemize}

\item $p \to \pi^0+e^+$ determine the grand unification (GUT) scale;

\item $p \to \pi^+ + \bar{\nu}$; $p \to K^+ + \bar{\nu}$; $n \to \pi^0 + \bar{\nu}$; $n \to K^0 + \bar{\nu}$ determine scale and constraint the parameters for the supersymmetry (SUSY) models;

\item $p \to \pi^0 + \mu^+$; $p \to K^0 + \mu^+$ determine the flavour structure of the models. 

\end{itemize}

Further, we restrict the discussion only to $p \to \pi^0+e^+$, $p \to \pi^+ + \bar{\nu}$, and $p \to K^+ + \bar{\nu}$ decay channels. The proton lifetime varies depending on the decay channel. Furthermore, as previously mentioned, lifetime limits have been established by multiple collaborations, with S-K providing the highest limit. Figure 1 reproduces the predictions for the proton decay lifetime for the considered channels from the paper of Perez et. al. in Ref.~\refcite{perez2019qcd}. The figure further illustrates that the proton lifetime determines the GUT scale.

\begin{figure*}[h!tb]
\centering
\vspace*{8pt}
\begin{subfigure}{0.3\textwidth}
    \includegraphics[width=1\linewidth]{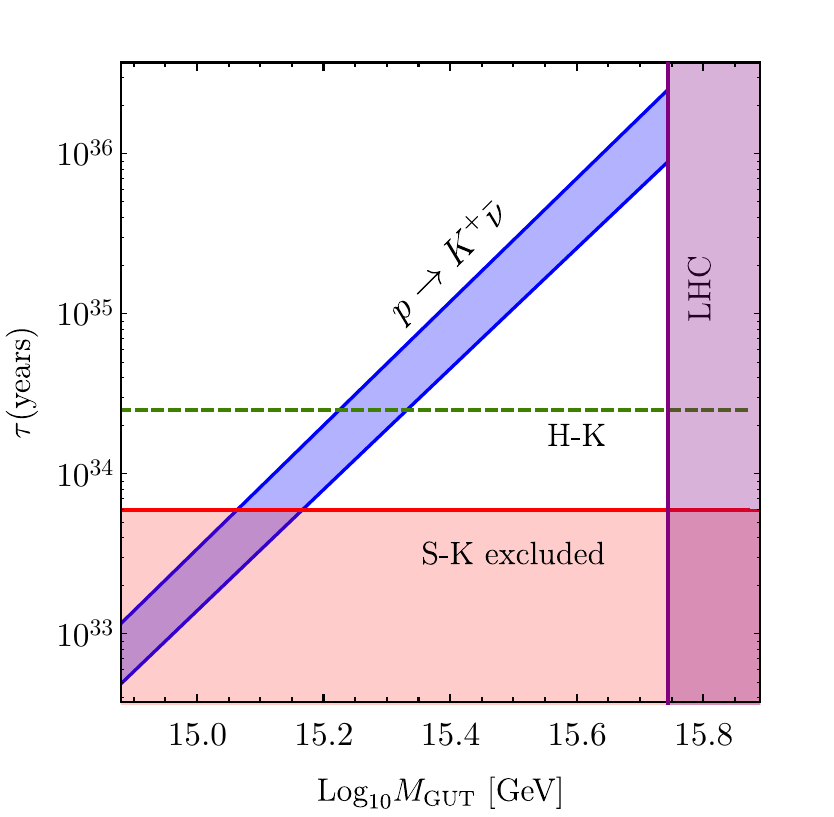}
\end{subfigure}
\begin{subfigure}{0.3\textwidth}
    \includegraphics[width=1\linewidth]{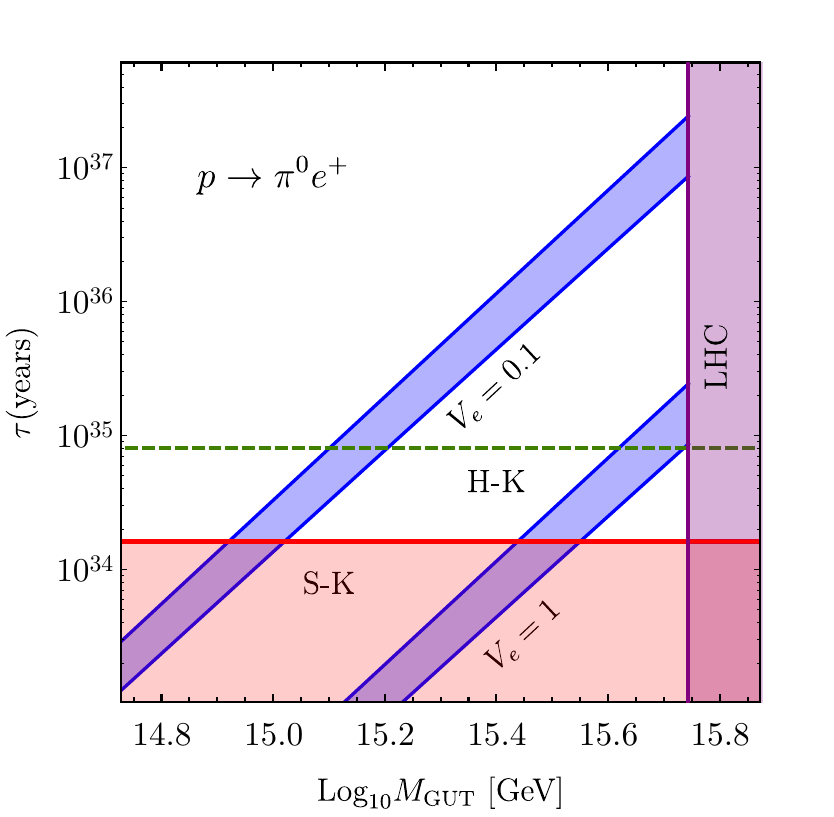}
\end{subfigure}  
\begin{subfigure}{0.3\textwidth}
    \includegraphics[width=1\linewidth]{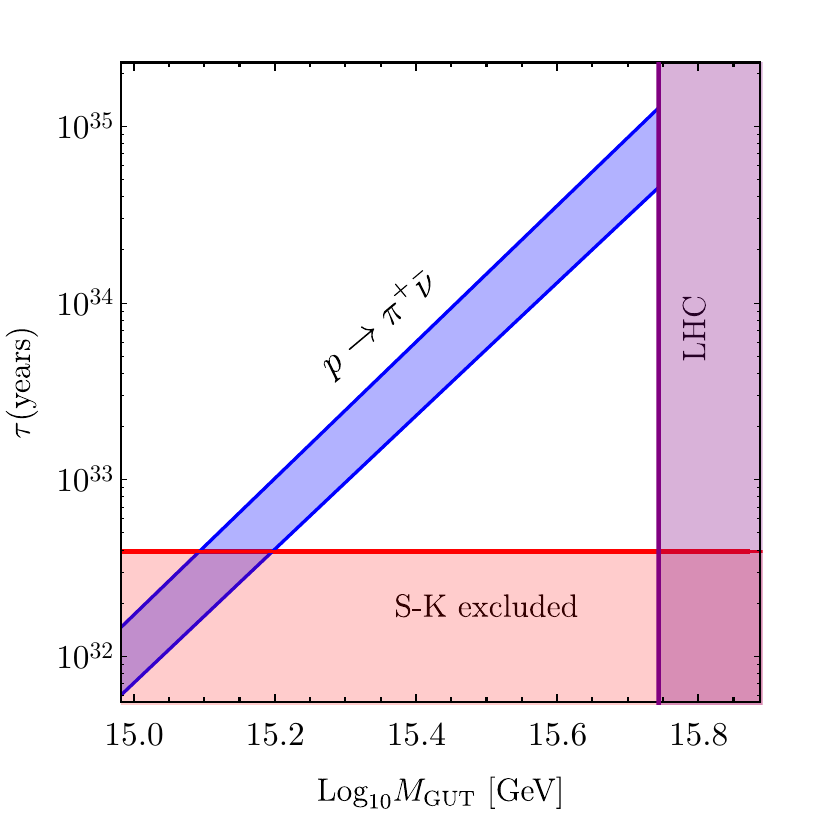}
\end{subfigure}
\caption{Predictions for the proton decay lifetime for the three channels \cite{perez2019qcd}. The width of the blue bands is determined by the allowed values of $\alpha_{GUT}$  scale. The red area indicates the excluded regions by different collaborations. The green line is the limit for lifetime imposed by the HK collaboration. $V_{e}$ refers to the magnitude of the mixing matrix.}
\label{proton_decay_lifetime}
\end{figure*}

\section{Proton decay experiments and detection techniques}

There are several detection techniques that are used by different experiments in proton decay searches \cite{OHLSSON2023116268}:

\begin{romanlist}[(ii)]
    \item Water Cerenkov Detectors. Cerenkov light is emitted along the surface of a cone of maximum half-angle 41\textdegree  (corresponding to $ \cos^{-1}\left(\frac{1}{n}\right) $) about the trajectory of a relativistic charged particle if $\beta > \frac{1}{n} = 0.75$. Older experiments for proton decay searches include IMB (Cleveland, Ohio), HPW (Park City, Utah), Kamiokande (Kamioka, Japan;1996, $\sim$22.5 kton) as well as next generation: S-K, Hyper-K.
    \item Tracking Calorimeters used the advantages of an iron calorimeter instrumented with layers of gas counters. Experiments: KGF (India), NUSEX (Mont Blanc tunnel), Frejus (in Frejus tunnel; used plastic flash tubes (25 mm$^2$), with Geiger tubes (225 mm$^2$)), Soudan (in Minnesota; gas ionization, time projection calorimeter).
    \item The next generation of experiments includes other techniques: Liquid noble gases as TPC: Ar (MicroBooNE, ProtoDUNE / DUNE), Xe (XENONnT, LUX-ZEPLIN (LZ)); Liquid scintillators (LENA).
\end{romanlist}

\section{Nuclear effects in proton decay}

The theoretical examination of the nuclear effects associated with nucleon decay in a nucleus has not been done sufficiently yet. In principle, some effects must be considered i.e., Fermi motion of the nucleons inside the nucleus, final state interaction of pion or kaon (elastic, inelastic, absorption), nucleon-nucleon correlation, virtual meson exchange processes. In this work, we analyzed only the first two effects mentioned before in three different materials, H$_2$O, LAr and LXe. A detailed understanding of the theories predicting proton or neutron decay is not essential for studying nuclear effects. \cite{Alvarez-Estrada:1986lmn}.

\subsection{Effects due to Fermi motion inside a nucleus}

We investigated only two of the hypothetical proton decay channels, i.e., $p\rightarrow \pi^++\bar{\nu}$ and $p\rightarrow K^++\bar{\nu}$. If the proton is considered to be at rest, then the kinematics of the resulting particles would be well defined. However, as we already mentioned, nucleons are moving inside the nucleus. This motion can be described by Global Fermi Gas Model, which states that the maximum momentum value of nucleons bound to a nucleus is the Fermi Momentum $k_F$ which depends on the nucleus. One approximation of the $k_F$ parameter based on the mass number $A$, can be seen in Ref.~\refcite{Bodek_2014}. In this sense, the values for the three nuclei we are studying are: $k_F(\mathrm{^{16}O})=228$ MeV, $k_F(\mathrm{^{40}Ar})=241$ MeV and $k_F(\mathrm{^{131}Xe})=245$ MeV. The momentum distribution for $k<k_F$ is described by:
\begin{equation}
    P(k)dk=\frac{4\pi k^2}{N}dk
    \label{eq1}
\end{equation}
where $k$ is the nucleon momentum and $N$ is defined as
\begin{equation}
    N=\frac{4}{3}\pi k_{F}^3
    \label{eq2}
\end{equation}

From Eq. \ref{eq1} and Eq. \ref{eq2} we obtained the momentum distribution $P(k)=\frac{k^3}{k_F^3}$ which can be seen in Fig. \ref{fig:Proton_FermiMomentum}.:


\begin{figure}[h!]
  \centering
  \includegraphics[width=0.7\textwidth]{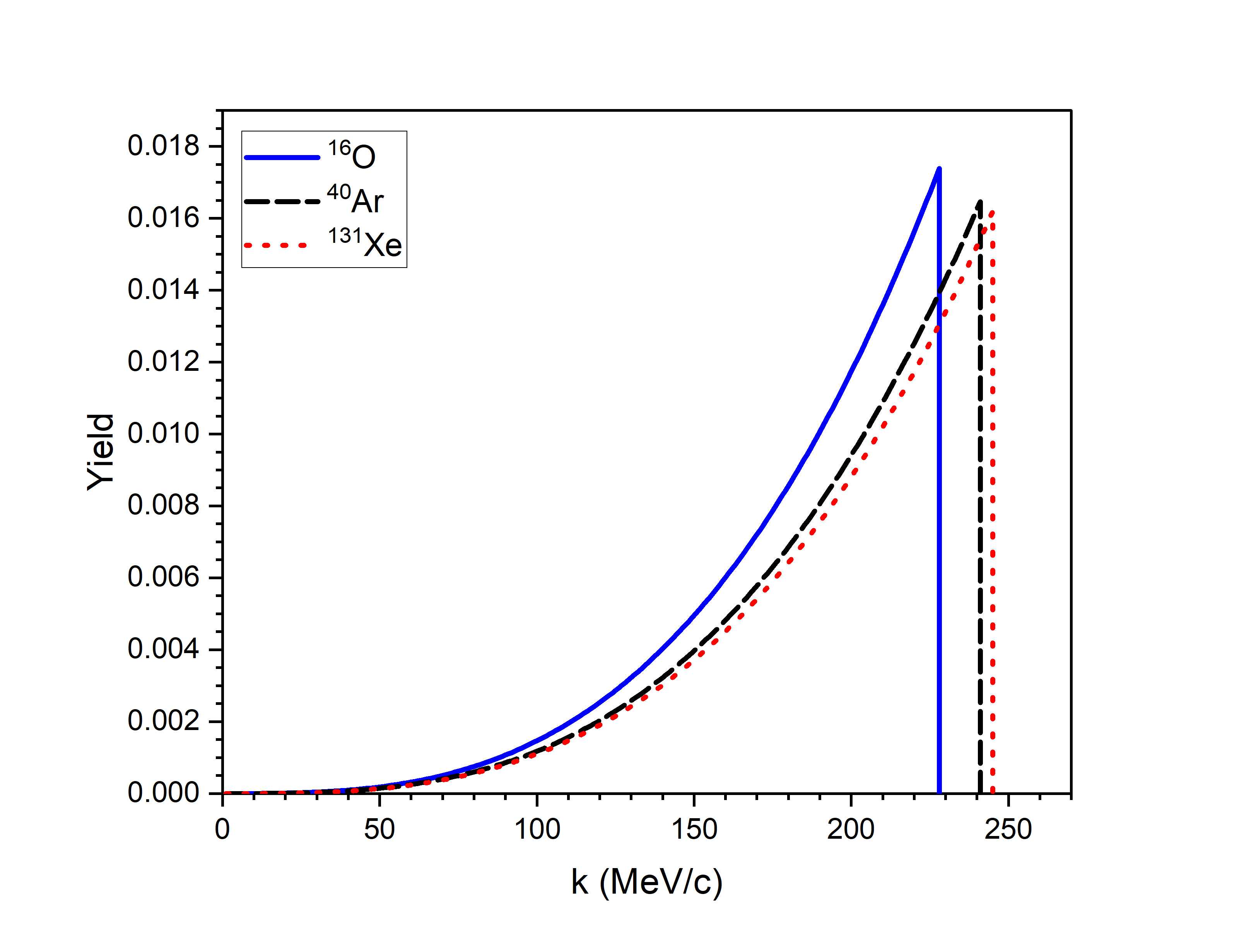}
  \caption{The nucleon momentum distribution based on the Global Fermi Gas Model for the three nuclei we are studying $^{16}$O, $^{40}$Ar and $^{131}$Xe. The distributions are normalized to unity.}
  \label{fig:Proton_FermiMomentum}
\end{figure} 

\begin{figure}[h!]
  \centering
  \includegraphics[width=.7\textwidth]{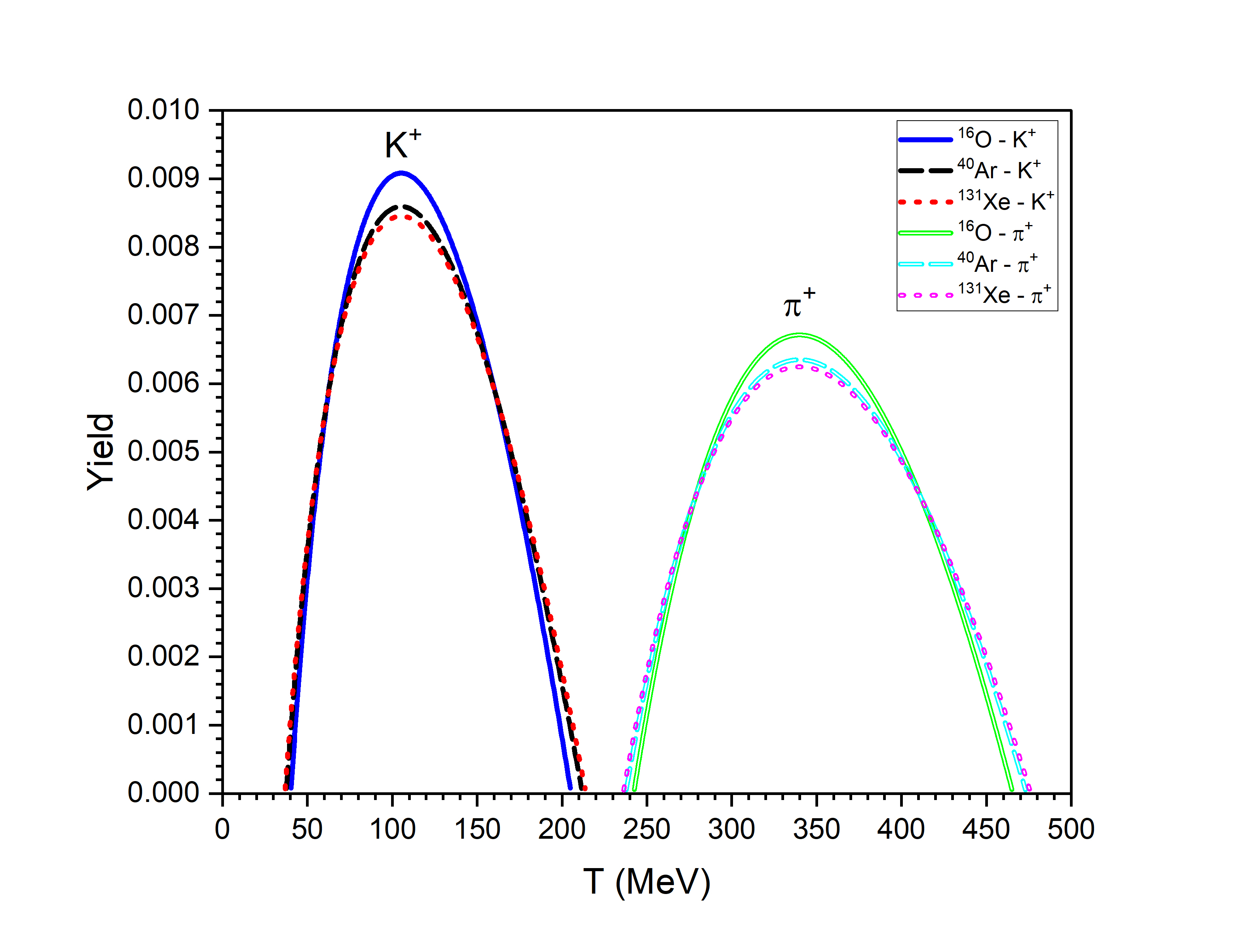}
  \caption{The kaon and pion energy distributions in a proton decay based on the Global Fermi Gas Model of the nucleus. The analysis was done for three nuclei, $^{16}$O, $^{40}$Ar and $^{131}$Xe. The distributions are normalized to unity.}
  \label{fig:kaon&pion}
\end{figure}

From the kinematics of the proton decay one can determine the limit values of the kinetic energies of the kaon or pion for each momentum of the decaying proton \cite{mcdonald2001off}:
\begin{equation}
    T_K = \frac{E_p}{m_p} \left(599 \, [\text{MeV}] \pm \frac{k_p}{E_p} \cdot 339.26 \, [\text{MeV}]\right) - m_K \, [\text{MeV}]
\end{equation}
\begin{equation}
    T_{\pi} = \frac{E_p}{m_p} \left(479.51 \, [\text{MeV}] \pm \frac{k_p}{E_p} \cdot 458.75 \, [\text{MeV}]\right) - m_{\pi} \, [\text{MeV}] 
\end{equation}
 where $E_p$, $m_p$ and $k_p$ are the total energy, the rest energy and the momentum of the proton and $m_K$, $m_\pi$ are the rest energies of kaon and pion. The energy distribution of the decay products (kaon or pion) is flat for a specific proton momentum. However, for different proton energy values, the distributions are not the same. Such a dependency can be described as \cite{mcdonald2001off}:
\begin{equation}
    Y(k_p)\propto \frac{1}{k_p}
    \label{eq6}
\end{equation}
In Fig. \ref{fig:kaon&pion} we show the obtained energy distributions for the two hypothetical proton decay modes, $p\rightarrow \pi^++\bar{\nu}$ and $p\rightarrow K^++\bar{\nu}$ considering the Global Fermi Gas Model and the kinematic limits detailed before. The most probable pion or kaon energies are the ones for the case of decay at rest. However, compared to the whole spectra, these energies are just a small fraction. For argon, the results seem to be in good agreement with Geant4 simulations, see Ref.~\refcite{Stefan:2008zi}. The spectral function approach has the potential to yield more realistic results (see Ref.~\refcite{Stefan:2008zi} and Ref.~\refcite{Alvioli:2010hv}).The distribution is broadened in the higher energy region. However, a more pronounced difference emerges when considering intranuclear cascade effects, regardless of whether the Fermi motion or spectral function frameworks are applied. 

\subsection{Effects due to pion-nucleus or kaon-nucleus interactions}

In this chapter, we studied the pion and kaon nuclear interactions in the nuclear medium i.e., inside the nucleus and in the active volume of the detectors (LAr, LXe, H$_2$O). In this sense, we provided results on cross-sections (total cross-sections for kaon, total and absorption cross-sections for pion), mean free paths, and probabilities for the particles to not interact. Because the data for these particles is poor (especially for pion), we will use different parametrizations to estimate the cross-sections.

\subsubsection{Pion interaction in nuclear medium.}
The total and elastic cross-sections for $\pi^+p$ and $\pi^-p$ scattering as a function of the total C.M. energy or as a function of the pion momentum in laboratory, put in evidence the character of interactions with production of $\Delta$ and N resonances. The productions of $\Delta$(1232) and $\Delta$(1950) are dominant in $\pi^+p$, while N(1520) and N(1680) are especially produced in $\pi^-p$, both can be observed in elastic and total cross sections. When these processes are investigated in nuclei, some peculiarities are put in evidence. In this case, free interactions are not accurate. Collective effects exist and modify the elementary process. The production of $\Delta$ resonance remains the dominant process in the kinetic energy region between 100 – 300 MeV of the pion. Other contributions are other resonances that are masked by the mentioned effects \cite{carroll1976pion,ashery1981true}. Here, it is important to mention that pion absorption is a process possible on at least two nucleons, and in the final state the pion disappears.

In the region of $\Delta$ resonance, the pion-nucleus total cross section is the result of approximately equal parts: elastic, inelastic (quasielastic $\pi$-N scattering), and absorption. For different contributions, the cross sections do not generally vary as A$^{2/3}$ thus the interactions are not produced only on the surface. In the particular case of absorption, the primary process is on 2 nucleons, but absorption involves more than 2 nucleons up to $>$70\% probability in conditions when the detailed mechanism is not well established.

The data for pion interactions in nuclear medium (inside the nucleus) is poor. We used a parametrization that was proposed by Cugnon et. al. in Ref.~\refcite{cugnon1996simple} which gives values of cross-sections of the pion with protons and neutrons:
\begin{equation}
    \\\sigma(\pi^+ p) = 3\sigma(\pi^+ n) = \frac{326.5}{1 + 4\left(\frac{\sqrt{s} - 1.215}{0.110}\right)^2} \cdot \frac{1}{1 + \left(\frac{0.18}{q}\right)^3}
\end{equation}
where $q$ is the C.M. momentum:
\begin{equation}
    \\q = \left[\frac{(s - (m_{\pi} + m_p)^2)(s - (m_{\pi} - m_p)^2)}{4s}\right]^{\frac{1}{2}} = \frac{m_p}{\sqrt{s}} p_{\text{lab}}
\end{equation}
and $s$ can be written as:
\begin{equation}
    \\\sqrt{s} = \sqrt{2m_p E_{\pi} + m_{\pi}^2 + m_p^2}
\end{equation}

The parametrization works only for pions which have the momentum lower than 0.7 GeV.
For a specific nucleus, since the $\sigma(\pi^+p)$ and $\sigma(\pi^+n)$ are different, we approximated the total cross section as
\begin{equation}
 \sigma(\pi^{+} \mathrm{^{A}_{Z}X}) = \frac{Z}{A} \cdot \sigma(\pi^+ p) + \frac{A-Z}{A} \cdot \sigma(\pi^+ n)
\end{equation}

\begin{figure*}[h!]
  \centering
  \begin{subfigure}{0.5\textwidth}
    \includegraphics[width=1\linewidth]{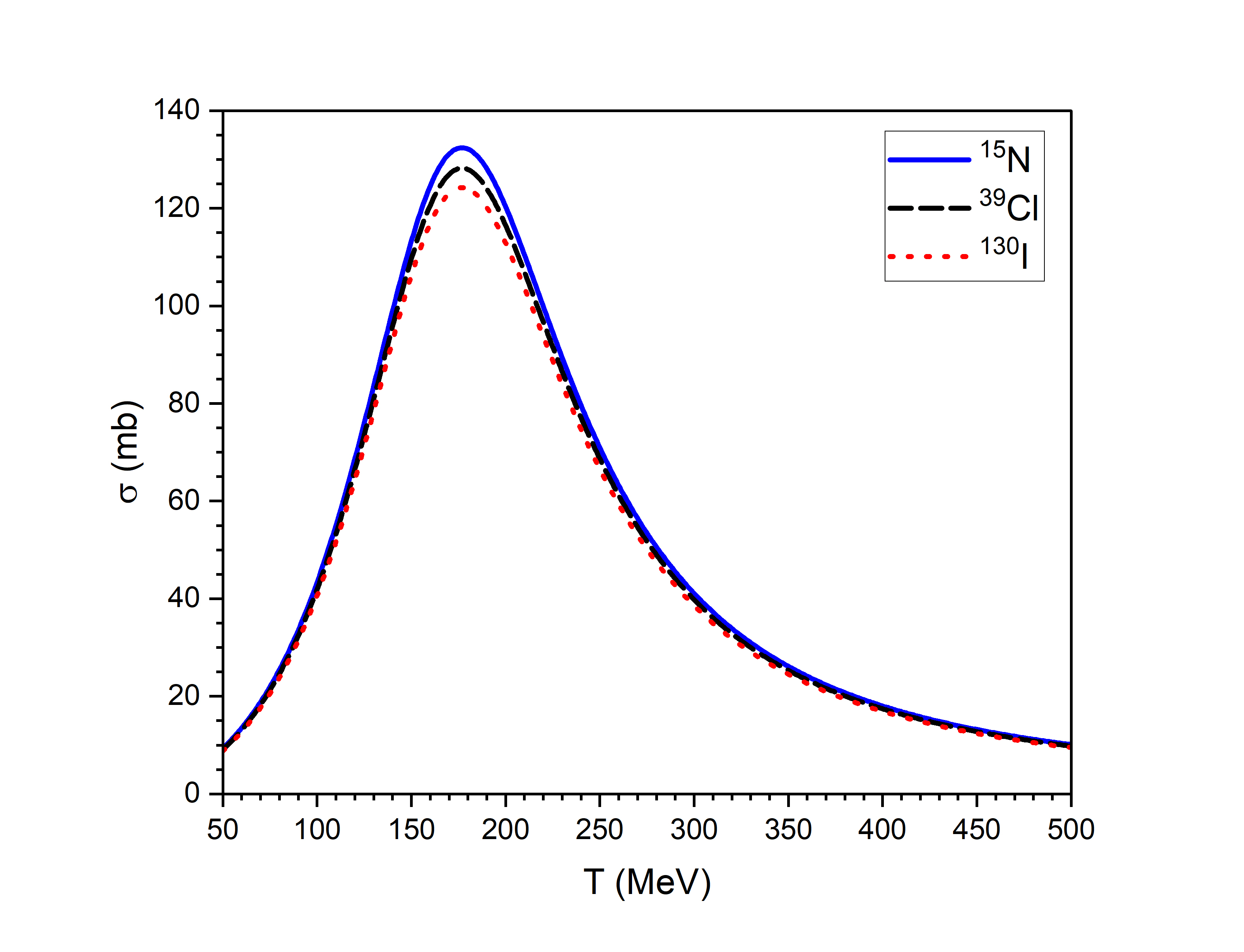}
    \caption{}
    \label{PionCS_nucl}
  \end{subfigure}%
  \begin{subfigure}{0.5\textwidth}
    \includegraphics[width=1\linewidth]{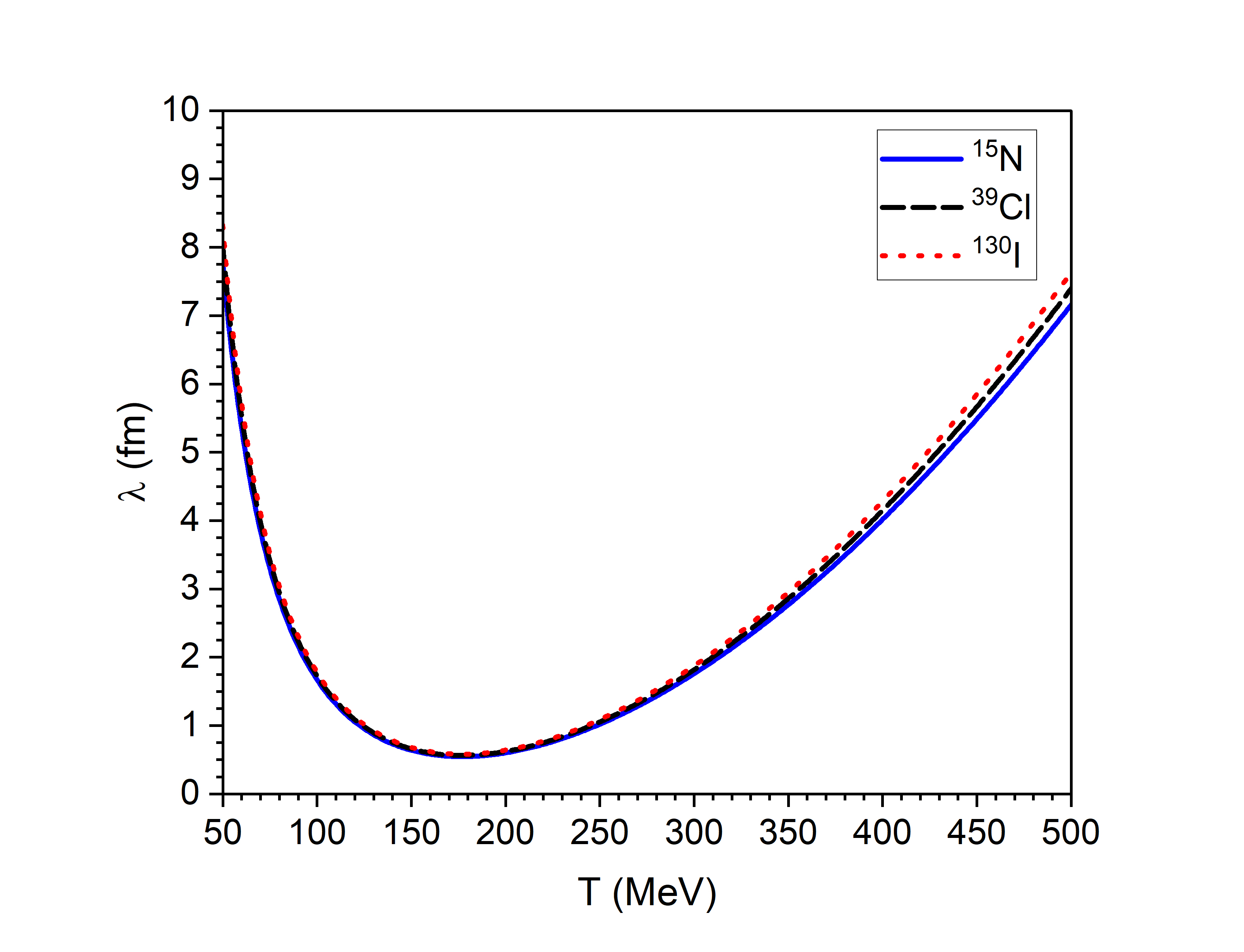}
    \caption{}
    \label{PionMFP_nucl}
  \end{subfigure}%
  \caption{(a) Pion total cross sections inside the nucleus. (b) Pion mean free path inside the nucleus. The results are given for three nuclei, $^{15}$N, $^{39}$Cl and $^{130}$I.}
      \label{pion_xs}
\end{figure*}

This pion cross section refers to an average pion-nucleon cross section. The notation suggests that these cross-sections are different for other nuclei based on the ratio of protons and neutrons. We consider that in the range of energy from Fig. \ref{fig:kaon&pion} the pion can interact directly with the nucleons (the associated de Broglie wavelength is comparable with the dimensions of the nucleon). The values obtained using this parametrization can be seen in Fig. \ref{PionCS_nucl}.

Knowing the cross sections and the nucleon density in nuclear medium described by the Liquid Drop Model ($n=\frac{3}{4\pi r_0^3}$ where $r_0=1.2$ fm), we calculated the pion mean free paths in three nuclei, $^{15}$N, $^{39}$Cl and $^{130}$I, 
\begin{equation}
    \\\sigma = \frac{1}{n \lambda} 
    \label{eq:11}
\end{equation}

The nuclei studied here are different from the ones from last chapter for a simple reason. Once the proton decays in a nucleus, that nucleus changes to a nucleus with $A_{new}=A-1$ and $Z_{new}=Z-1$. That means $^{16}$O goes into $^{15}$N, $^{40}$Ar goes into $^{39}$Cl and $^{131}$Xe goes into $^{130}$I, Fig. \ref{PionMFP_nucl}.

\begin{figure}[h!]
  \centering
  \includegraphics[width=.7\textwidth]{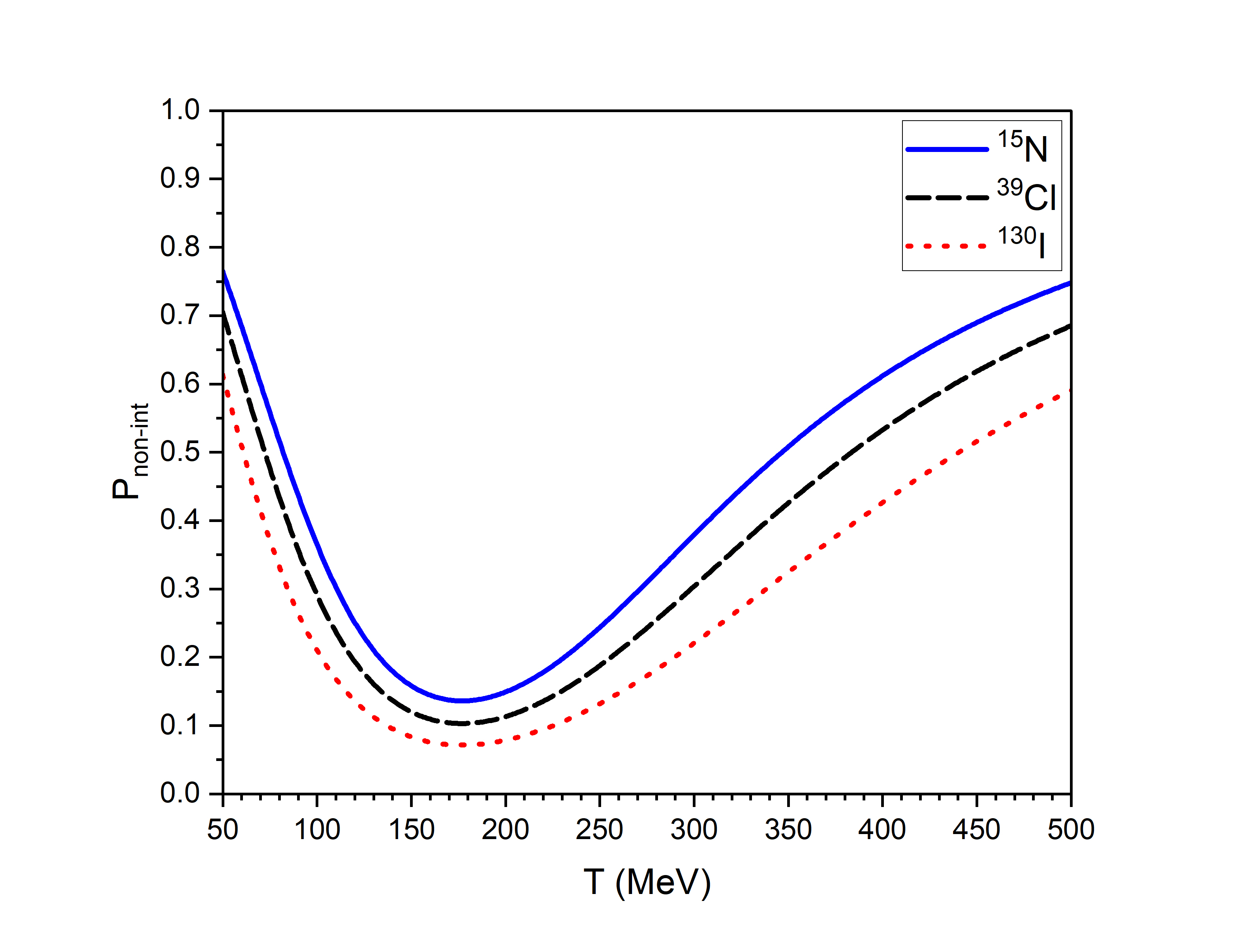}
  \caption{The non-interaction probabilities of the pion inside the residual nucleus in a proton decay for three nuclei, $^{15}$N, $^{39}$Cl and $^{130}$I.}
  \label{fig:PionNonIntProb_nucleus}
\end{figure}

Furthermore, of interest is the probability for the pion to not interact inside the residual nucleus. In the case when the pion leaves the nucleus without any interaction with the nucleons, one would expect the kinematic limits obtained in Fig. \ref{fig:kaon&pion}. However, if the pion interacts with one or more nucleons inside the residual nucleus, then the kinematics would drastically change from the ones provided in Fig. \ref{fig:kaon&pion} with the possibility of the pion vanishing in the case of absorption. This probability can be calculated as it follows \cite{nishimura1983nuclear}:
\begin{equation}
    \\P_0(A) = \frac{\int d^3 x \, e^{-L/\lambda}}{\int d^3 x} = \frac{3}{8} \left(\frac{\lambda}{R_A}\right)^3 \left[2\left(\frac{R_A}{\lambda}\right)^2 - 1 + \left(1 + 2\frac{R_A}{\lambda}\right) e^{\frac{-2R_A}{\lambda}}\right]
    \label{eq:12}
\end{equation}
where $L$ is the distance traveled by the pion inside the nucleus (which depends on where in the nucleus the decay happened and in which direction the pion was emitted) and $R_A$ is the nuclear radius of the residual nucleus using the Liquid Drop Model ($R_A=r_0A^{1/3}$). The dependence of this probability with the initial energy of the pion can be seen in Fig. \ref{fig:PionNonIntProb_nucleus}.

\subsubsection{Pion interaction in the active volume.}

In this chapter, we analysed the nuclear interactions of the pions with the nuclei of the detector medium. We have used the parametrization suggested by Ashery et. al. in Ref.~\refcite{ashery1986pion} in order to estimate the cross-sections of pions with $^{1}$H, $^{16}$O, $^{40}$Ar and $^{131}$Xe:
\begin{equation}
    \\\sigma = \sigma_0 A^N
    \label{eq:13}
\end{equation}
where $\sigma_0$ and $N$ are variables that depend on the energy and $A$ is the mass number.

\begin{figure*}[h!]
  \centering
  \begin{subfigure}{0.45\textwidth}
    \includegraphics[width=\linewidth]{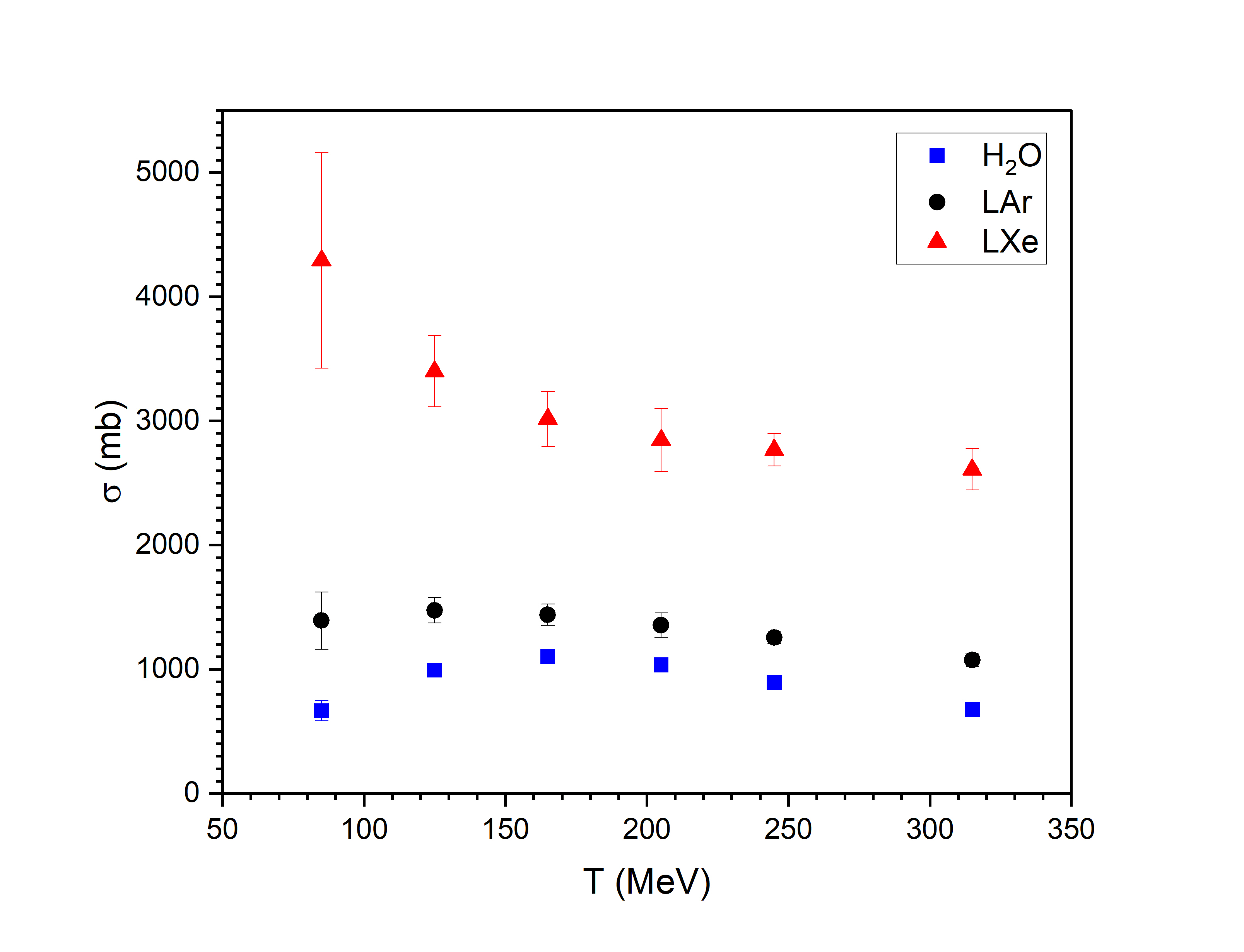}
    \caption{Total pion-nucleus cross section inside the active volume}
    \label{pion_cs_volume}
  \end{subfigure}\hfill
  \begin{subfigure}{0.45\textwidth}
    \includegraphics[width=\linewidth]{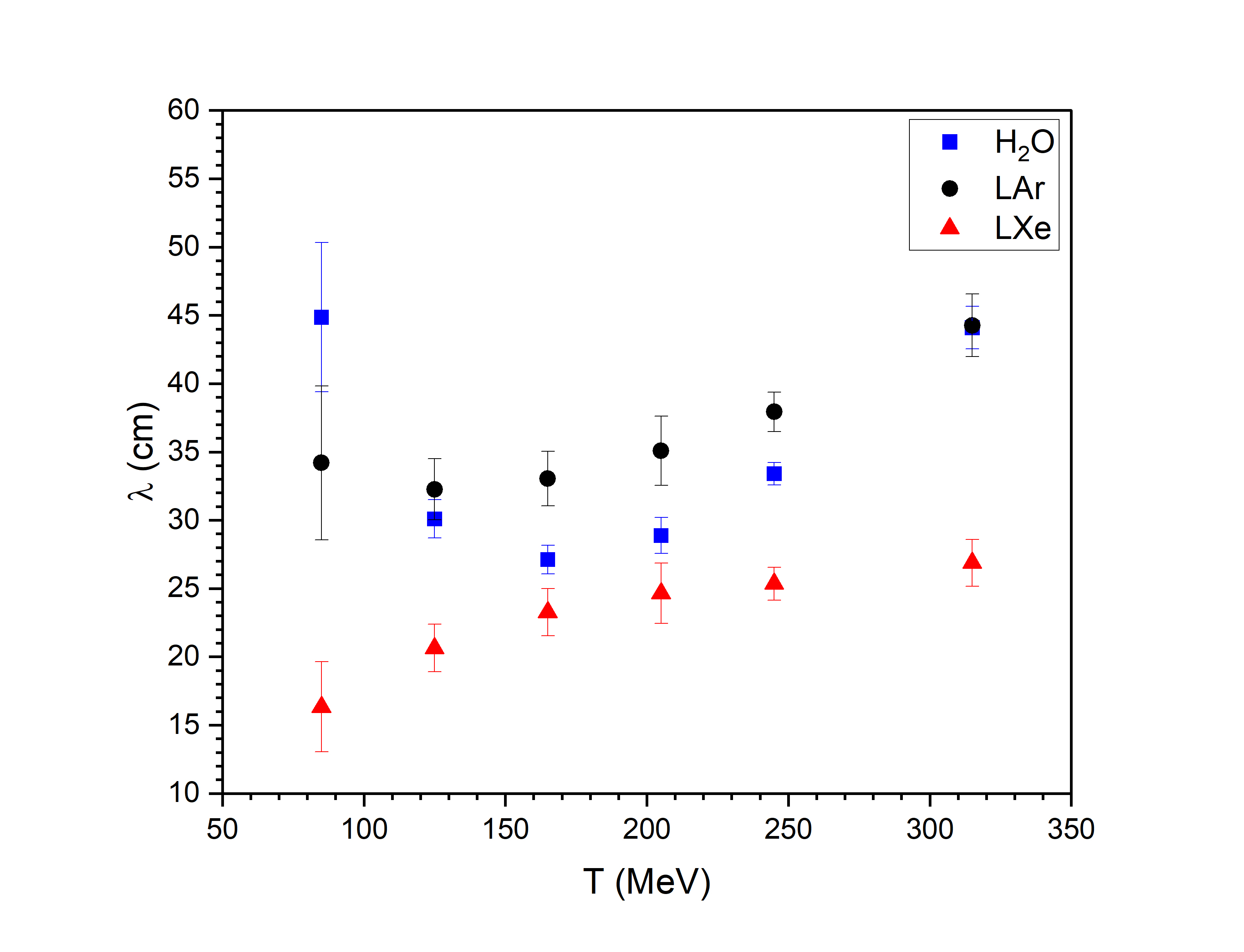}
    \caption{Total pion-nucleus mean free path inside the active volume}
    \label{pion_mfp_volume}
  \end{subfigure}\\[1ex]
  \begin{subfigure}{0.45\textwidth}
    \includegraphics[width=\linewidth]{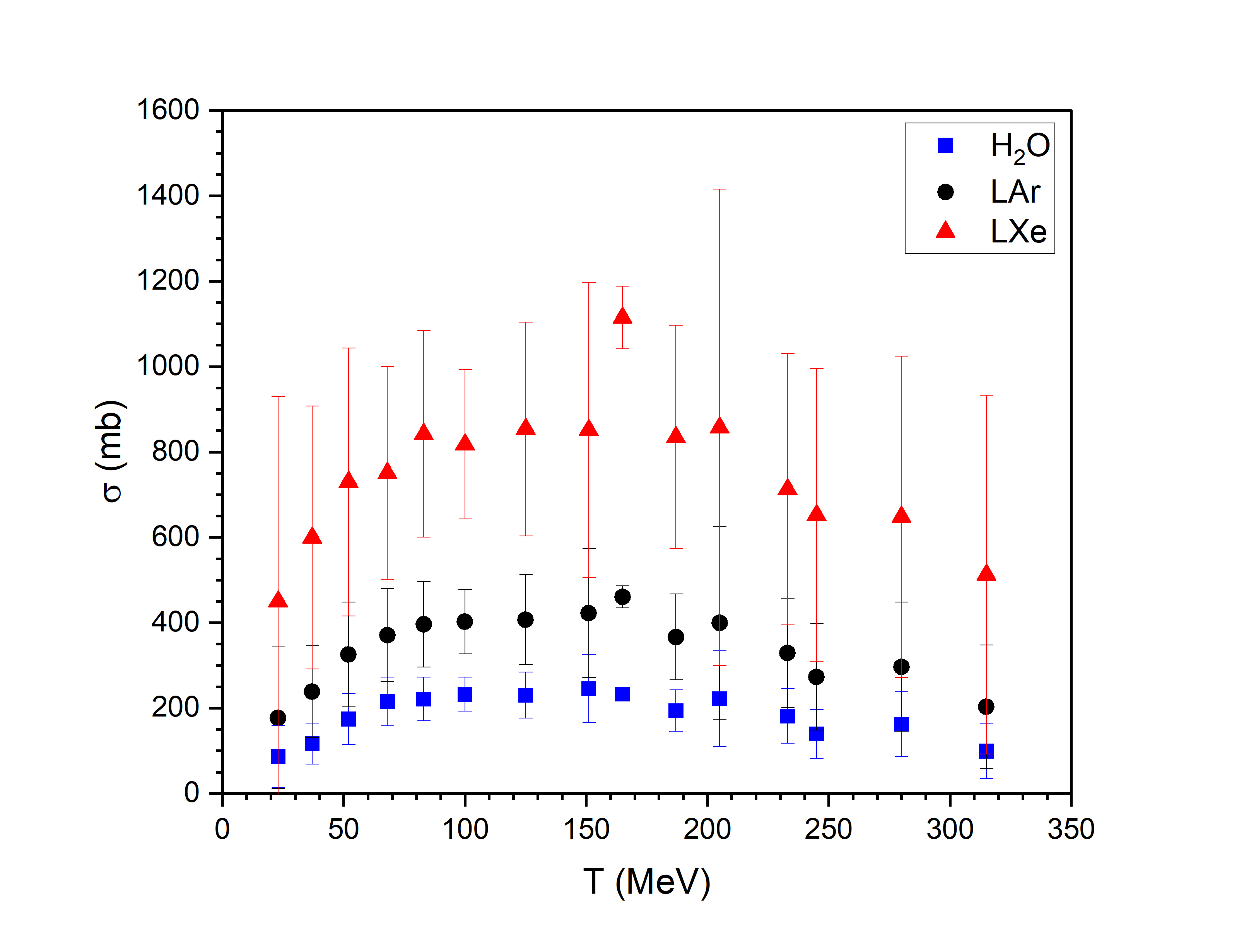}
    \caption{Absorption pion-nucleus cross section inside the active volume}
    \label{pion_abscs_volume}
  \end{subfigure}\hfill
  \begin{subfigure}{0.45\textwidth}
    \includegraphics[width=\linewidth]{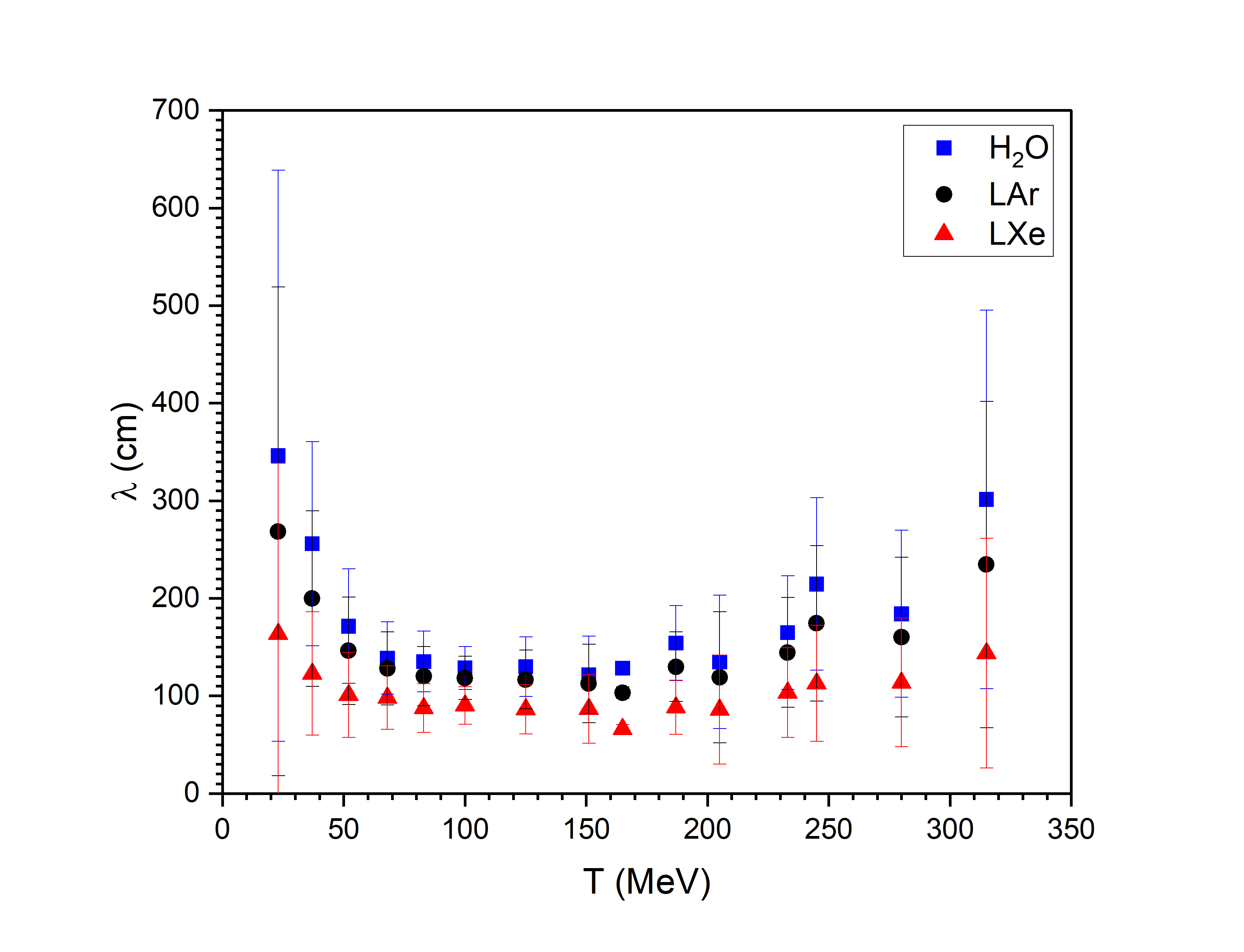}
    \caption{Absorption pion-nucleus mean free path inside the active volume}
    \label{pion_absmfp_volume}
  \end{subfigure}
  \caption{The results are provided for three different mediums, H$_2$O, LAr, and LXe.}
\end{figure*}

Using this parametrization and the data from Ref.~\refcite{ashery1981true} and Ref.~\refcite{nakai1980measurements} for the following nuclei: $^{7}$Li, $^{12}$C, $^{27}$Al, $^{48}$T, $^{56}$Fe, $^{64}$Cu, $^{92}$Nb, $^{119}$Sn, $^{197}$Au, $^{209}$Bi we estimated the total and absorption pion cross sections for H$_2$O, $^{40}$Ar, $^{131}$Xe for different energies, Fig. \ref{pion_cs_volume} and Fig. \ref{pion_abscs_volume}. In order to estimate the total cross-section for the H$_2$O medium, we used an approximation. We considered that $\sigma (\pi^+\,\mathrm{H_2O})=\sigma(\pi^{+}\,\mathrm{^{16}O})+2\cdot \sigma(\pi^{+}\,\mathrm{^{1}H})$. In the case of absorption, since this process can take place only on at least two nucleons, the contribution from the hydrogen atoms is neglected for the H$_2$O medium.
The total and absorption pion mean free paths are calculated using Eq. \ref{eq:13} (See Figures \ref{pion_mfp_volume} and \ref{pion_absmfp_volume}).
Even though we calculated these quantities based on their experimental uncertainty, after using the parametrization described by Eq. \ref{eq:13}, all these uncertainties are partially lost. The error bars in our data refer mainly to uncertainties arising from the parametrization.

\subsubsection{Kaon interaction in nuclear medium.}
The experimental data for the kaon-nucleon cross-section at low energies is limited, and there is an absence of direct measurements for kaon-neutron interactions. Therefore, we adopted the following parametrization to estimate the kaon-neutron cross sections: $\sigma(K^+ p) = \sigma_1$ and $\sigma(K^+ n) = \frac{1}{2}(\sigma_1 + \sigma_0)$ \cite{Castelli:1974ym}, where $\sigma_1$ and $\sigma_0$ correspond to the cross sections of the two isospin states, $I=0$ and $I=1$ for which we have experimental data Ref.~\refcite{Cameron:1974xx} and Ref.~\refcite{Carroll:1973ux}. In the energy range of 20 - 150 MeV for the $\sigma_0$ we have extrapolated linearly the data between 150 - 290 MeV. A simple approximation of the average kaon-nucleon cross section inside different nuclei could be:
\begin{equation}
    \sigma(K^{+} \mathrm{{^A_Z}X}) = \frac{Z}{A} \cdot \sigma(K^+ p) + \frac{A-Z}{A} \cdot \sigma(K^+ n)
    \label{eq:15-1}
\end{equation}

\begin{figure}[h!]
  \centering
  \begin{subfigure}{0.45\textwidth}
    \includegraphics[width=1\linewidth]{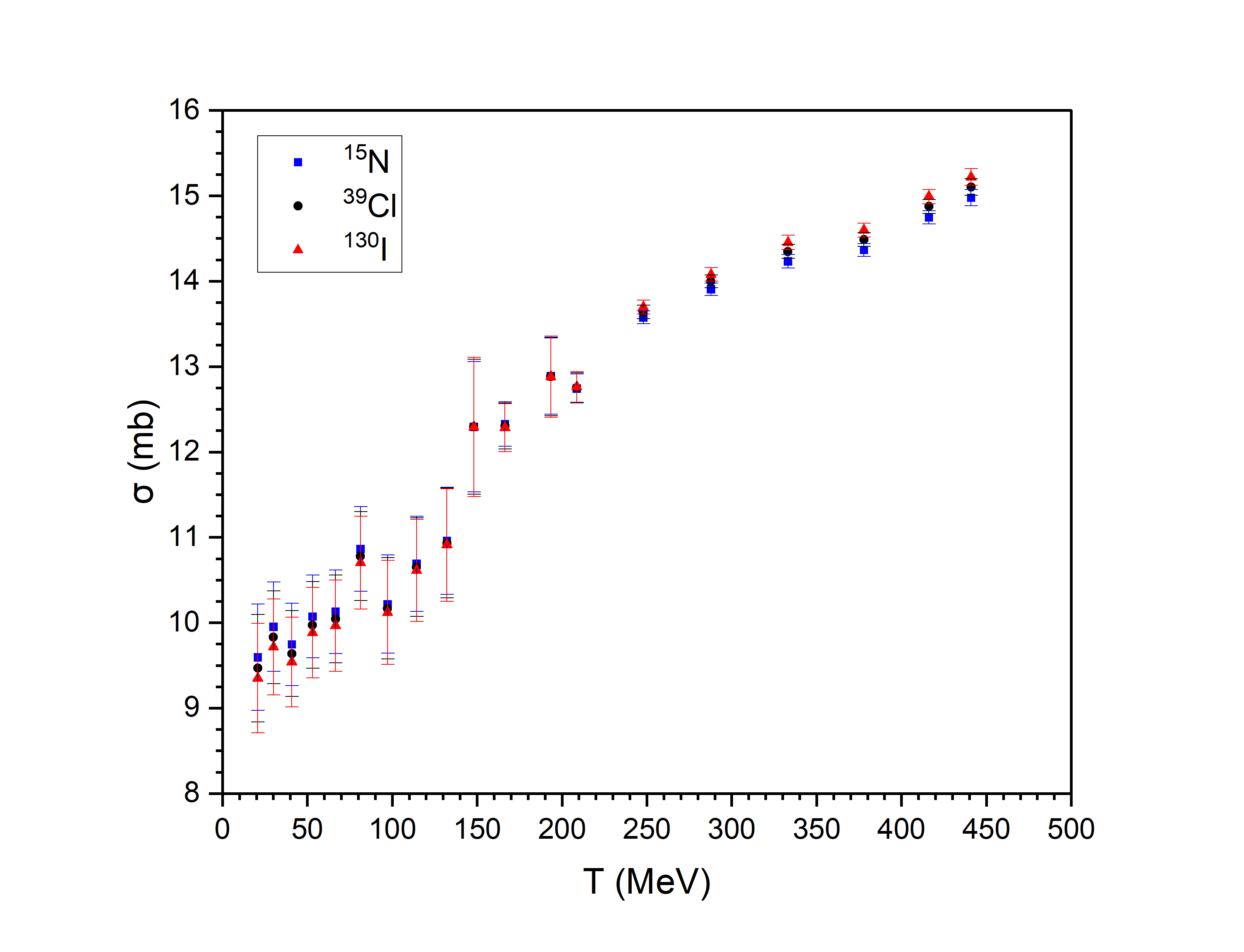}
    \caption{}
    \label{kaon_cs_nucleus}
  \end{subfigure}
  \begin{subfigure}{0.45\textwidth}
    \includegraphics[width=1\linewidth]{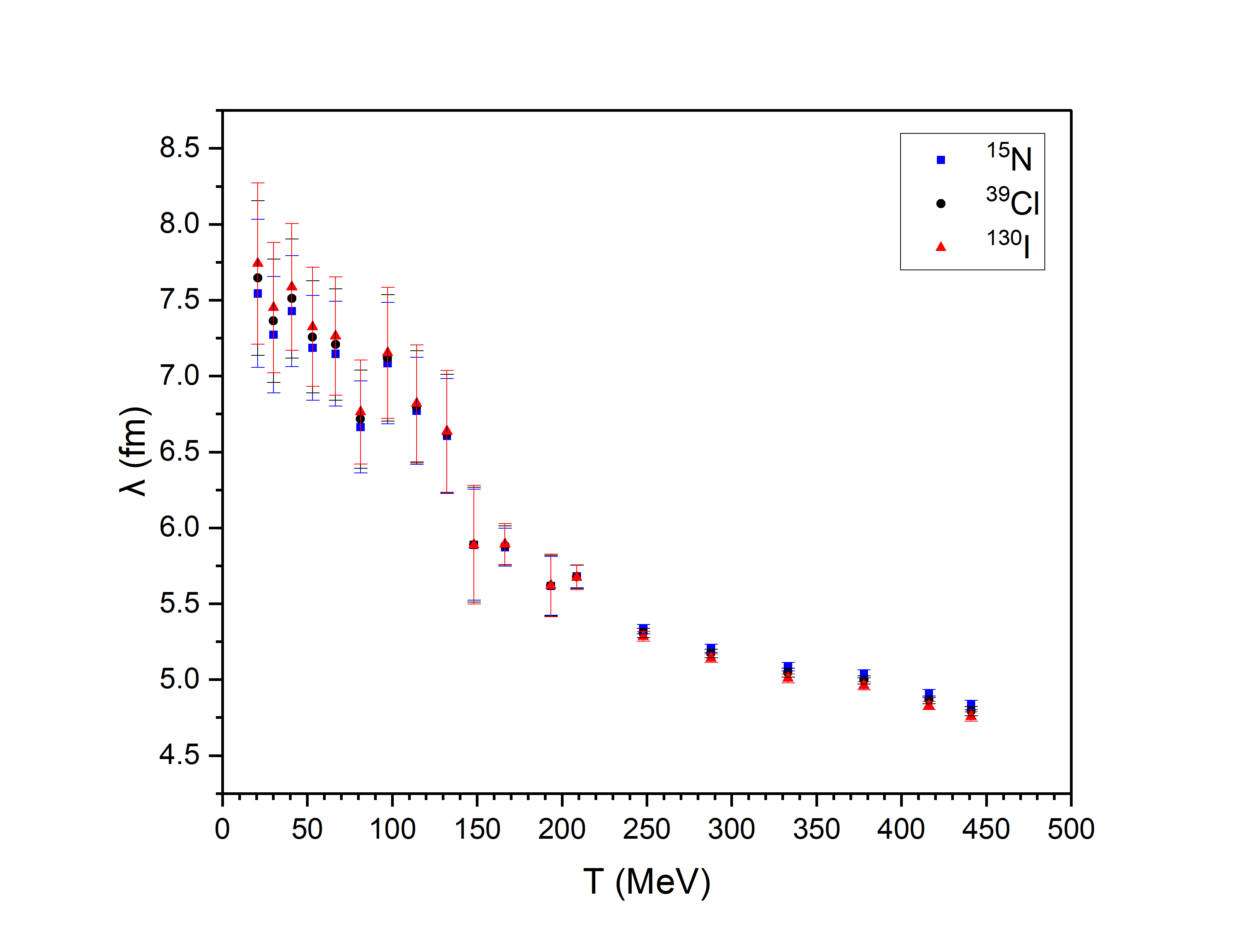}
    \caption{}
    \label{kaon_mfp_nucleus}
  \end{subfigure}  
  \caption{a) Kaon cross sections inside the residual nucleus. b) Kaon mean free paths inside the residual nucleus.}
      \label{kaon_xs}
\end{figure}

The cross sections were calculated for three nuclei: $^{15}$N, $^{39}$Cl, and $^{130}$I, corresponding to the residual nuclei resulted from proton decay in each of the three previously discussed materials (Fig. \ref{kaon_cs_nucleus}). The hydrogen atom ($^1$H) was not considered in these calculations, as proton decay in $^1$H results in the disappearance of the nucleus.

\begin{figure}[h!]
  \centering
  \includegraphics[width=.7\textwidth]{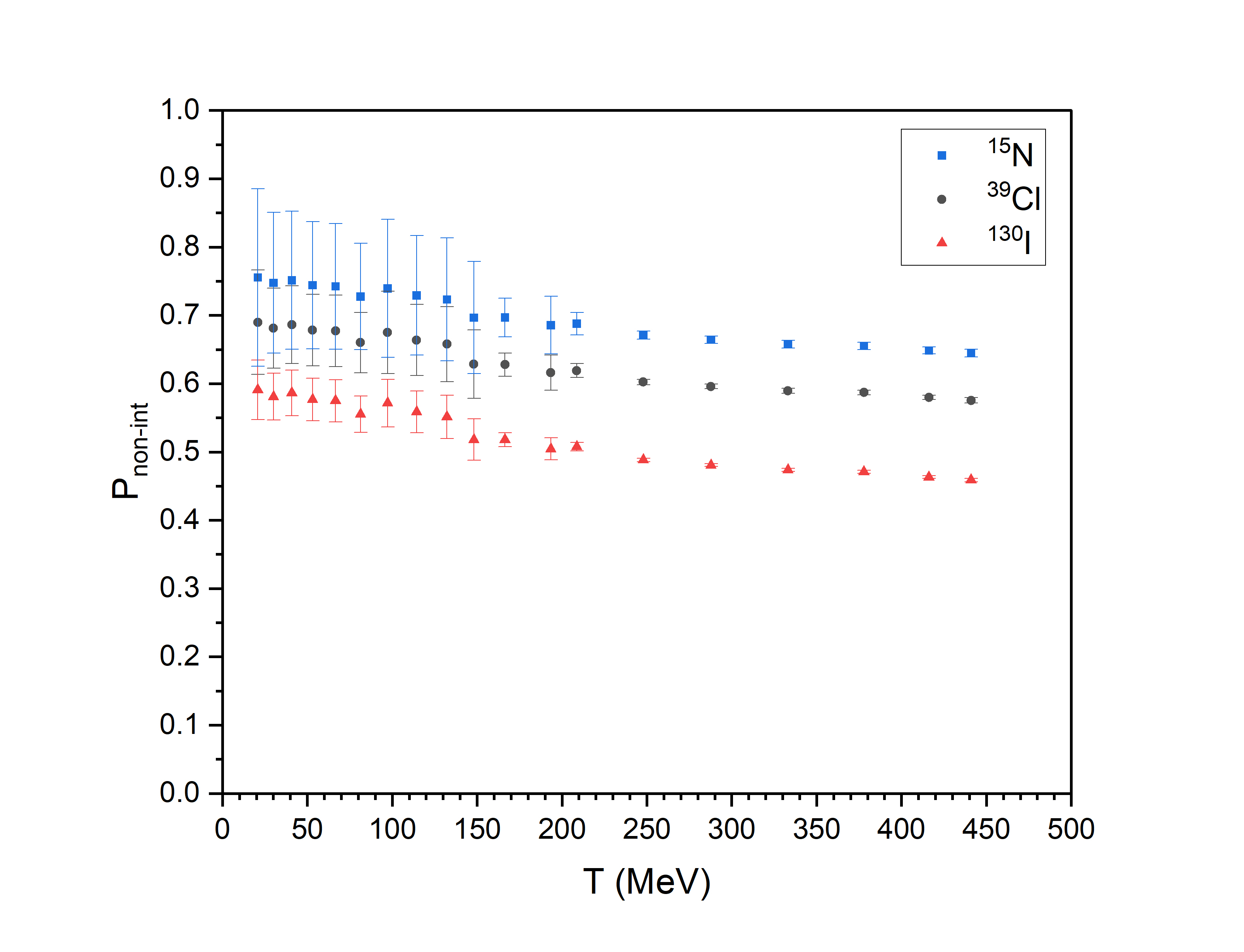}
  \caption{Probability of the kaons to not interact with the nucleons inside the residual nuclei, $^{15}$N, $^{39}$Cl and $^{130}$I.}
  \label{fig:kaon_nonint_nucleus}
\end{figure}

Knowing the nucleon density for each nuclei discussed, we calculated the mean free paths of the kaons inside the residual nuclei (Fig. \ref{kaon_mfp_nucleus}). Lastly, we computed the non-interaction probabilities using Eq. \ref{eq:12} (See Fig. \ref{fig:kaon_nonint_nucleus}).

\subsubsection{Kaon interaction in the active volume.}
The interaction of kaons with the nuclei of the active volume is also of interest. However, unlike pions, positive kaons can not be absorbed.  The problem of  extrapolation of the data from nucleons to nuclei was investigated by Weiss et. al. in Ref.~\refcite{weiss1994measurement} and they give the following approximation $\sigma(K^+nucleus)=A\sigma(K^+nucleon)$. In this paper we are using different cross sections for neutron and proton so the kaon-nucleus cross section can be written as:
\begin{equation}
    \\\sigma (K^+ \text{nucleus})=Z\sigma(K^+p)+(A-Z)\sigma(K^+n)
\end{equation}

For the kaon-water cross sections we used the approximation, $\sigma (K^+\,\mathrm{H_2O})=\sigma(K^{+}\,\mathrm{^{16}O})+2\cdot \sigma(K^{+}\,\mathrm{^{1}H})$.

\begin{figure*}[h!]
  \centering
  \begin{subfigure}{0.45\textwidth}
    \includegraphics[width=1\linewidth]{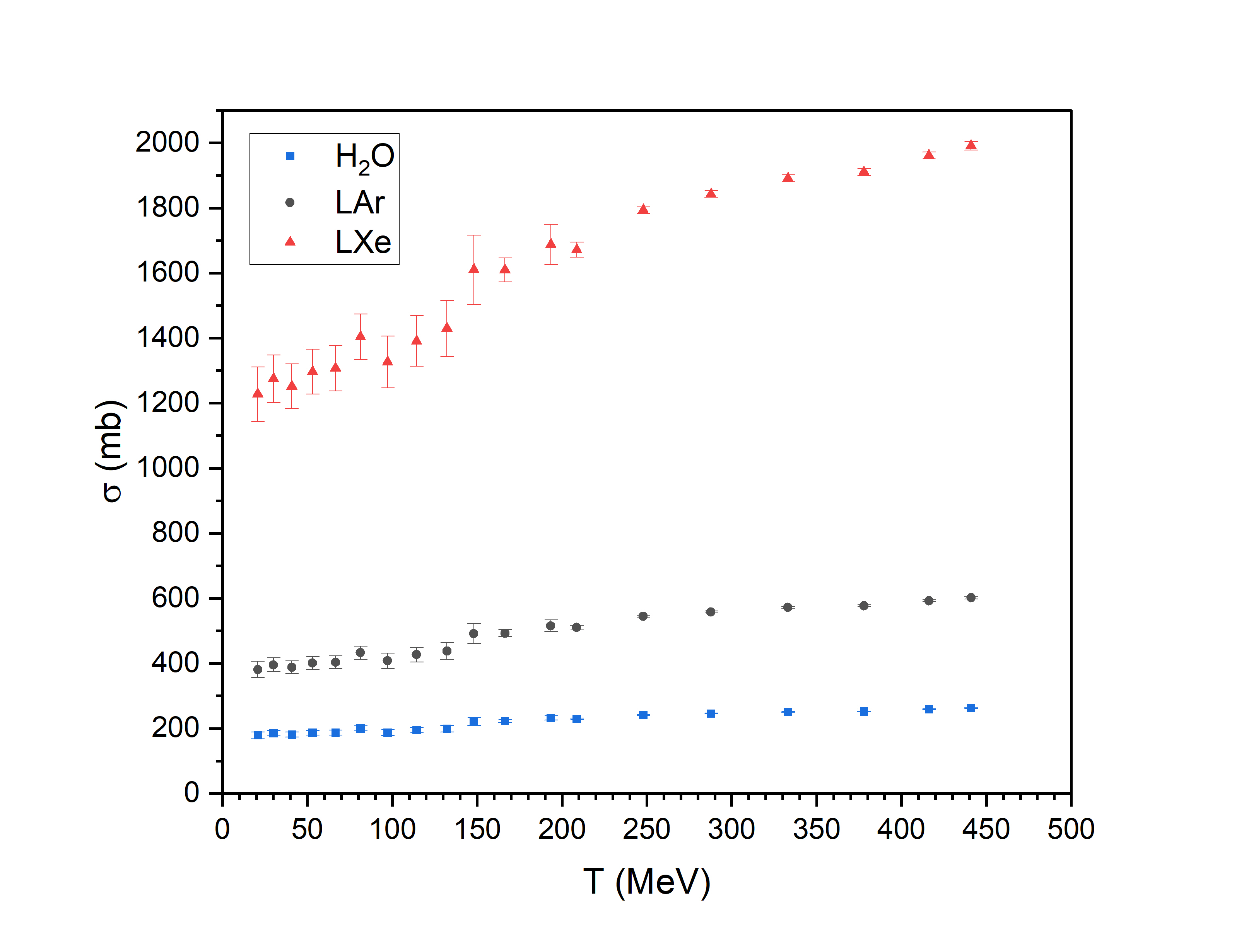}
    \caption{}
    \label{kaon_cs_volume}
  \end{subfigure}
  \begin{subfigure}{0.45\textwidth}
  \includegraphics[width=1.\linewidth]{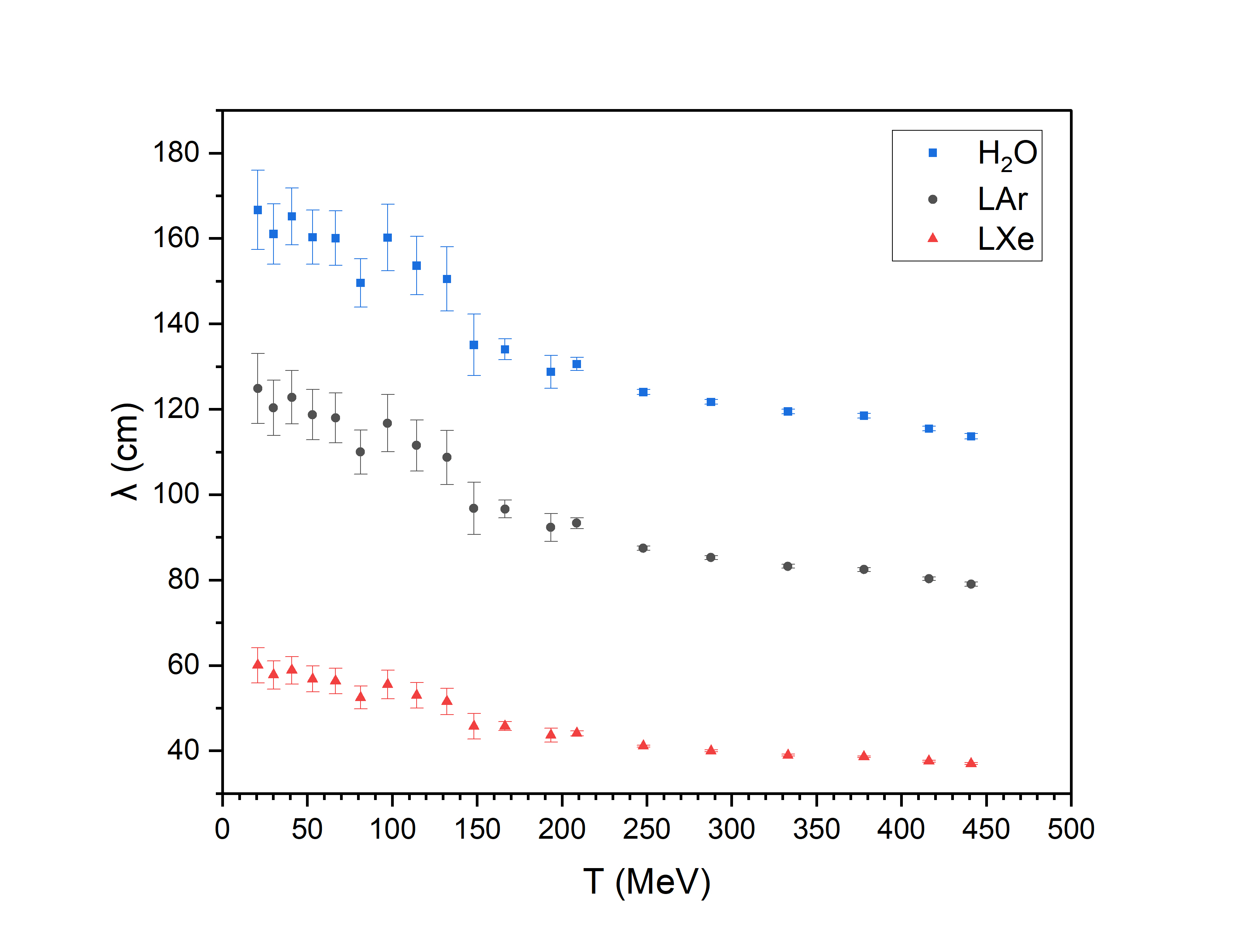}
    \caption{}
    \label{kaon_mfp_volume}
  \end{subfigure}%
  \caption{a) Kaon total cross sections inside the detector active volume. b) Kaon mean free paths inside the detector active volume.}
      \label{}
\end{figure*}

The mean free paths in the active volume are obtained from Eq. \ref{eq:11}. The dependence of kaon cross sections and mean free paths in the active volume with the energy is shown in Fig. \ref{kaon_cs_volume} and \ref{kaon_mfp_volume}.

\section{Summary}

In this paper, we analyzed two nuclear effects that must be considered in proton decay, Fermi motion of nucleons bound to a nucleus, final state interactions. These effects drastically change the kinematics of the decay from the case of an isolated back to back decay. We studied two hypothetical decay modes, i.e. $p\rightarrow \pi^++\bar{\nu}$ and $p\rightarrow K^++\bar{\nu}$ in three different mediums that are used for proton decay searches, H$_2$O, LAr and LXe. Using the Global Fermi Gas Model of the nucleus, we calculated the energy limits and the distribution of the kaon and pion in proton decay. Moreover, becuase the cross section data for kaons and pions with nuclei is poor, we used different parametrizations to estimate the values for the cases of interest in our study. In this sense, we provided values for pion and kaon total cross sections and mean free paths (inside the residual nucleus and also with the nuclei of the medium). Because the pion can be absorbed in its interaction with a nucleus if the energy constraints are overcome, we studied separately the pion-nucleus absorption cross-sections and mean free paths.

\section*{Acknowledgments}

This work is supported by contract no. 04/2022, Programme 5, Module 5.2 CERN-RO and CERN-RO/CDI/2024\_001.

\bibliographystyle{ws-mpla}
\bibliography{ws-mpla}

\end{document}